\documentstyle[psfig,amssymb,amsfonts]{aa}

\newcommand{\nc}{\newcommand}

\nc{\Teff}{$T_{\rm eff}$\,}
\nc{\Porb}{$P_{\rm orb}$\,}

\def\3he{$^3$He}
\def\12sur13{$^{12}$C/$^{13}$C}
\nc{\logg}{log~$g$~}
\nc{\kms}{\,${\rm km.s}^{-1}$}
\nc{\Ciso}{${\rm ^{13}C}$}
\nc{\C}{${\rm ^{12}C}$}

\begin{document}
\thesaurus{     
            08.01.1;  % Stars: abundances
            08.05.3;  % Stars: evolution
            08.12.1;  % Stars: late-type
            08.18.1.  % Stars: rotation.
}

\title{ Rotation and lithium in single giant stars
\thanks{Based on observations collected at the
Observatoire de Haute -- Provence (France) and at the European
Southern Observatory, La Silla  (Chile).}}

\author{
         J. R. De Medeiros,
         J. D. do Nascimento Jr,
         S. Sankarankutty,
         J. M. Costa,
         M. R. G. Maia  
                 }

\offprints{renan@dfte.ufrn.br}

\institute{Departamento de F\'{\i}sica, Universidade Federal do Rio
             Grande do Norte, 59072-970  Natal,  R.N., Brazil }

\date{Received July  2000, Accepted XX 2000}

\maketitle
\markboth{De Medeiros, do Nascimento, Sankarankutty, Costa, and  Maia:
Rotation and lithium in evolved giants stars}{Rotation and lithium 
in evolved giants stars} 

\begin{abstract}
{\sf  In the present work, we study the link between rotation and
lithium abundance in giant stars of luminosity class III, on the basis of
a large sample of 309 single stars of spectral type F, G and K.
We have found a trend for a link between the discontinuity in rotation at the 
spectral
type G0III and the behavior of lithium abundances around the same spectral 
type. 
The present work also shows that giant stars
presenting the  highest lithium contents, typically stars earlier than G0III,
are those with the highest rotation rates, pointing for a dependence of lithium 
content on 
rotation, as observed for other luminosity classes. Giant stars later than G0III 
present,
as a rule, the lowest rotation rates and lithium  contents. A large spread of 
about five magnitudes 
in lithium abundance is observed for the  slow rotators. Finally, single giant 
stars  with masses $ 1.5 < M/M_{\odot}\leq 2.5 $ show a clearest trend for a
correlation 
between rotational 
velocity and lithium abundance.  
}

\end{abstract}

\section{Introduction}

 One of the best known properties of the late-type evolved stars is that
their rotational velocity and lithium content decrease with age.
Nevertheless, the root cause of this property, as well as the relationship
 between rotation and lithium abundance, are not yet completely established.
 Which physical processes control the behavior of rotation and lithium
once stars evolve along the giant branch? In particular, how does rotation
 affect lithium dilution? How does magnetic braking affect rotation and
lithium content in evolved stars? Solid answers to these
 questions have been hampered by the difficulties in the measurement of
rotation rates for cool stars as well as by the paucity of lithium
abundance measurements for large and complete sample of evolved stars.
However, over the last 10 years it has become possible to measure
rotational velocities of evolved stars with a precision better than
1.0 km\,s$^{-1}$ (e.g.: Gray 1989; De Medeiros \&  Mayor 1989, 1999).
In addition, large observational surveys of the lithium line  6\,707.81 \AA,
with high precision, were carried out (e.g.: Brown {\it et al.} 1989;
Wallerstein {\it et al.} 1994; Balachandran 1990; Randich {\it et al.} 1999;
L\'ebre {\it et al.} 1998). As a result, some very interesting new features
on the behavior of rotation and lithium content in evolved
stars are emerging.

A dramatic drop in the rotation of subgiant and giant stars occurs,
respectively, near the spectral types F8IV and G0III (Gray \& Nagar
1985; Gray 1989; De Medeiros \& Mayor 1989, 1991). For the
more luminous classes, namely bright giant and Ib supergiant stars, De
Medeiros \& Mayor (1989, 1991) observed a sudden decrease in their
rotation near the spectral types F9II and F9Ib, respectively. On the other hand, 
the giant stars, to the left of the rotational discontinuity, namely in the F 
spectral region,
 present a wide range of rotational
velocity values, from a few km\,s$^{-1}$ to about  one hundred times the solar
rotation rate. To the right of this discontinuity, namely along the G and
K spectral regions, the large majority of stars show low rotation. 
Along this spectral region only binary systems presenting orbital periods 
shorter than
about 250 days and circular or nearly circular orbits and a dozen of apparently
single  giant stars present enhanced rotation.
Should one expect some kind of link between the rotational discontinuity
and the
   distribution of lithium abundances?
A gradual decrease of surface lithium abundance is expected
 once stars evolve along the giant branch. At the end of the
main-sequence, theory predicts that lithium is confined to the outermost
 regions of the star in a thin convective layer. Once the star evolves
up the giant branch, the convective envelope expands towards the stellar
 interior and the convective mixture of the outer material rich in
lithium with deeper and Li-free material leads to the depletion of this
fragile element (e.g.: Iben 1967a,b). Lithium is destroyed when the
convective envelope with Li-rich material reaches stellar inner regions
with temperatures  higher than about $2.5\times 10^{6}$ K. Following the
initial study by Bonsack (1959), different studies have attempted to
 analyse  the observational behavior of the lithium abundance along the
giant branch. The abundance of lithium in F and early-G giants has been
 investigated by Wallerstein (1966), Alschuler (1975) and Wallerstein
{\it et al.} (1994). These works have shown a steady decline in lithium
abundances from F5III to F8III, a wide spread around the spectral type
G0III and low values up to G5III. Wallerstein {\it et al.} (1994) have  
analysed
the behavior of lithium abundances of F2 -- G5 giants looking for the main
features of lithium in the Hertzsprung gap. These authors have found
that rapidly rotating giants, namely stars with $v\sin i>50$ km\,s$^{-1}$, 
located
 in the color interval from  $(B-V)$ = 0.40 to
 $(B-V) = 0.70$ present lithium abundances close to the presumed
primordial value $\log n(Li) = 3.0$. For  slow rotators,
$v\sin i<50$ km\,s$^{-1}$, a drop in lithium abundance appears in the
color interval  $(B-V) = 0.45$ to  $(B-V) = 0.60$, i.e. stars
with  $(B-V)>0.45$ show reduced lithium abundances in spite of
their early spectral types. Luck (1977), Lambert {\it et al.} (1988) and
Luck and Lambert (1982) have determined lithium abundances for normal
G, K and M-type field giants. From these works, strong evidences
have emerged that the content of lithium in this spectral region is primarily
controlled by the stellar mass. On the basis of a very large survey of about 644 
G
and K-type giants Brown {\it et al.} (1989) have found that a small
percentage of late-type giant stars have lithium content far in excess
of the standard predictions, a few of them approaching the primordial
value. The lithium content distribution of the remaining stars in such
survey shows that giants only very rarely are in agreement with standard
 first dredge-up predictions, in the sense that their lithium content
falls below the theoretical predictions. De Medeiros {\it et al.} (1996a) 
have
shown that, except for a few  chromospherically active giants, Li-rich evolved 
stars
show normal rotational velocity with respect to the typical lithium-normal 
evolved
stars of the same spectral type. Recently, a study of lithium abundance for 
population I 
subgiants
carried out by Lebre at al. (1999) have shown a sudden decrease in the lithium 
abundance
around Teff =  5\,600 K, corresponding to a spectral type near G2IV. Such a 
feature occurs
slightly later than the rotational discontinuity at F8IV. Furthermore, some 
dependence of lithium
abundance on rotation, in the sense that fastest 
rotators
show enhanced
lithium content, was observed in different classes of luminosity. This 
dependence, for 
example, was found for population I subgiant (Randich {\it et al.} 1999, De 
Medeiros {\it et al.}  1997, do 
Nascimento {\it et al.} 2000) and for unevolved late-type stars in young 
clusters 
(Garcia Lopez {\it et al.} 1994, Randich {\it et al.} 1998).

In spite of the important results obtained from the works
discussed above, up to date there have been no studies on the 
link between rotation and lithium content along the giant branch.
In the present work we investigate specifically the relationship between 
rotational
velocity and lithium abundance for giant stars on the basis of a
large and homogeneous data sample for which we have now precise
rotational velocities obtained with the CORAVEL spectrometer.

\section{Observational data}

The data sample selected for the present investigation has as its main
characteristics the high precision of the rotational velocity and
lithium abundance, as well as the large size of the sample. The
rotational velocity measurements were taken from the
Catalog of Rotational and Radial Velocities for Evolved Stars by the
De Medeiros \& Mayor (1999). By using the CORAVEL spectrometer
(Baranne {\it et al.} 1979) these authors have measured rotational
 velocities for a large sample of about 2\,000 evolved stars. For
giant stars of luminosity class III, in particular, these authors
 have shown that CORAVEL $v\sin i$ values present an
uncertainty of about 1.0 km\,s$^{-1}$ for stars with
 $v\sin i$ lower than about 30.0 km\,s$^{-1}$. For higher
 rotators, the estimations indicate an uncertainty of about 10\% .
For a complete discussion on the observational procedure,
 calibration and error analysis the reader is  referred to De Medeiros
and Mayor (1999).
\begin{figure}
\vspace{.2in}
\centerline{\psfig{figure=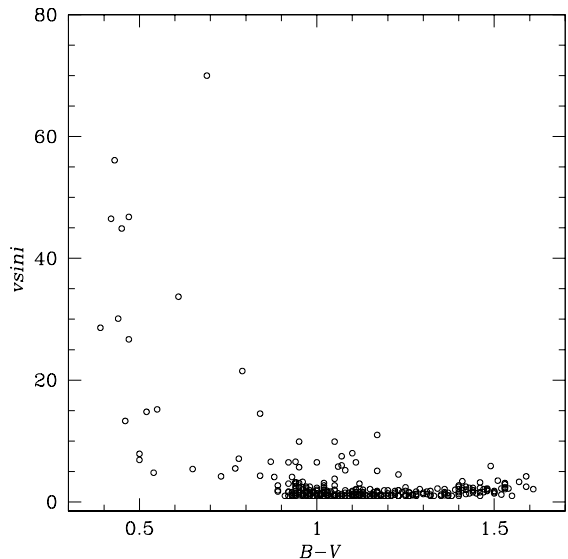,width=3.1truein,height=3.1
truein}
\hskip 0.1in}
\caption{The rotational velocity $v\sin i$ of single giant stars as a function 
of color index $(B-V)$. 
}
\label{Vsini_BV_Sing}
\end{figure}
Lithium abundances were taken from three different sources:
16 F and early-G stars from Wallerstein {\it et al.} (1994), 7 F stars from
Balachandran (1990) and 286 late-G and K type giants from Brown {\it et al.}
(1989), all of them with a $v\sin i$ given by De Medeiros \&
Mayor (1999). Because lithium abundances come from different authors,
a comparison for those stars in common in the given sources would be
important.  For four stars observed by Wallerstein {\it et al.}
(1994) and Balachandran (1990) a least-square solution yields a linear 
correlation coefficient of about 0.91 and standard deviation about 0.27, 
indicating for an excellent agreement between the lithium abundance values obtained by these 
authors. Brown {\it et al.} (1989) and Wallerstein {\it et al.} (1994) present only two stars 
in common.
Let us underline here the uncertainties in lithium abundances estimated by the
different authors mentioned above.  Wallerstein {\it et al.} (1994) estimate an 
uncertainty of
$\pm 0.1$ dex in $[Li/Fe]$ for stars presenting low rotation and weak Li
lines and
$\pm 0.3$ dex for rotating stars with strong Li lines, but for rotating
stars with  $(B-V)<0.5$ such uncertainty may  rise by 0.1 or 0.2
dex. Balachandran (1990) estimates uncertainties for $\log n(Li)$ to be
$\pm 0.02$ dex at large equivalent widths and $\pm 0.07$ dex at small
equivalent widths, while for rapidly rotating stars the error in
lithium abundances is probably $\pm 0.1$ dex. Finally, Brown {\it et al.}
(1989) have estimated an uncertainty of about $\pm 0.2$ dex. Moreover,
Wallerstein {\it et al.} (1994) obtained spectra with a signal-to-noise ratio of
about 200, Balachandran (1990) pointed out a signal-to-noise ratio of about
100 for the spectra, whereas all spectra obtained by Brown {\it et al.}
(1989) showed a signal-to-noise ratio of at least 150. In fact, the detailed
error analysis made by these authors indicate that their lithium
abundance measurements have the same high quality.
The entire sample with rotational velocities and lithium abundances is listed
in electronic table at CDS (http://cdsweb.u-strasbg.fr/)

\begin{figure}
\vspace{.2in}
\centerline{\psfig{figure=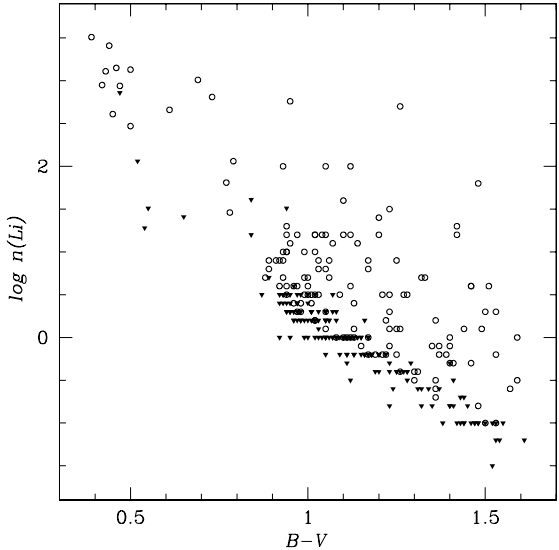,width=3.1truein,height=3.1truein}
\hskip 0.1in}
\caption{Lithium abundance as a function of color index $(B-V)$ for single giant stars. Upper limits for 
$\log n(Li)$ are indicated by triangles.
}
\label{ALi_BV_Sing}
\end{figure}

\section{Results and discussion}

The rotational velocity $v\sin i$ as a function
 of the color index $(B-V)$ is plotted in  Figure \ref{Vsini_BV_Sing}, where one 
sees
clearly the sudden decline in rotation near  $(B-V)$ = 0.70,
corresponding to the spectral type G0III (De Medeiros \& Mayor 1989, 1991, Gray 
1989). This cutoff 
in the distribution of the rotational velocity
results from a mixing in age and masses associated with the rapid evolution of 
giant
stars into the Hertzsprung gap (De Medeiros \& Mayor 1991). To the left of 
this
cutoff one sees a wide range of rotational velocity values, which seems to 
reflect the
distribution of rotation of the progenitors of giant stars. To the right of the 
cutoff, as pointed
out by De Medeiros \& Mayor (1991), single giants with high rotation
are unusual. Only a dozen of single G and K  single giants, in addition to 
synchronized binary 
systems, present enhanced rotation. As shown
by De Medeiros {\it et al.} (1996b), to the right of the cutoff rotation 
decreases 
smoothly from about 6.0
km\,s$^{-1}$ at G1III, to about 3.0 km\,s$^{-1}$ at G5III and to about 2.0 
km\,s$^{-1}$ along the 
spectral region from G8III to K7III.

\begin{figure}
\vspace{.2in}
                 
\centerline{\psfig{figure=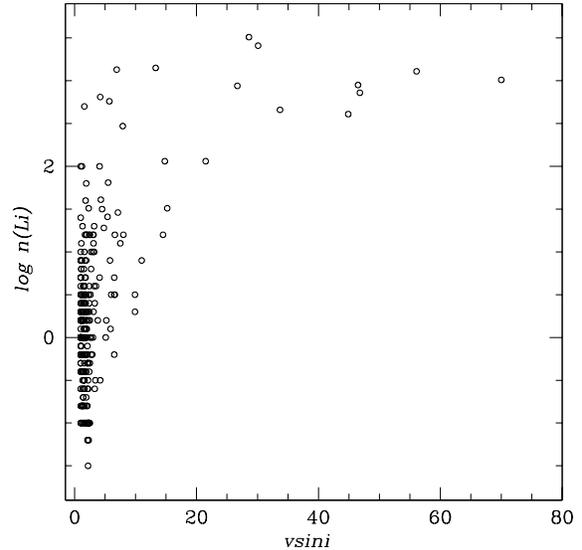,width=3.1truein,height=3.1truein}
\hskip 0.1in}
\caption{Lithium abundance as a
function of rotational velocity $v\sin i$ for single giant stars.
}
\label{Vsini_ALi_Single}
\end{figure}

\begin{figure}
\vspace{.2in}
      
\centerline{\psfig{figure=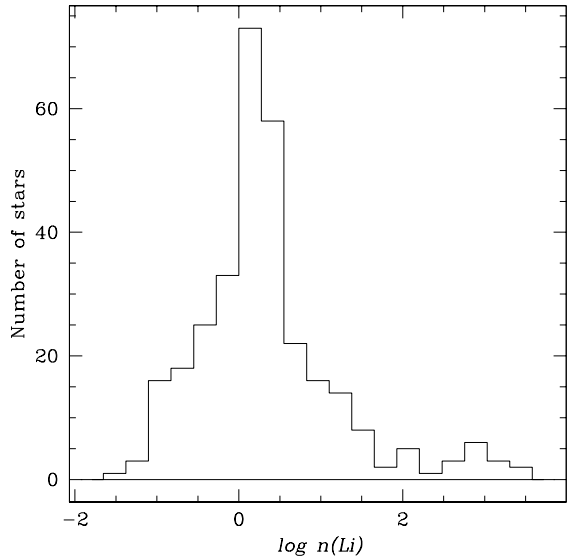,width=3.1truein,height=3.1truein}
\hskip 0.1in}
\caption{A histogram plot of the distribution of the lithium abundances for 
single giant stars.}
\label{hist_ALi_Single}
\end{figure}

The well established gradual decline of lithium abundances as a function
of the color index  $(B-V)$ for single giant stars is illustrated in
Figure \ref{ALi_BV_Sing}.
Let us recall that the $(B-V)$ interval represented in
that figure covers the spectral
range from F2III to K5III. Such a decline is interpreted as an evidence
of convective mixing. The convective zone begins to grow towards the
stellar interior as the star evolves from the F to the
G spectral region and finally deepens rapidly as the star ascends the first
giant branch, so that the thin surface layer containing lithium is mixed
 with the larger inner Li-free stellar material. Lithium is then burned
 when the convective envelope drags the mixed material into regions with
 temperatures of about $2\times 10^{6}K$. However, such observed decline in
 lithium abundance is not yet fully explained by the standard stellar
evolution theory. Along the giant branch the lithium abundance falls
below the standard predictions. In fact, standard theory shows
a factor of dilution of about 40 to 60 for 1.0 M$_{\odot }$ and
2.0 M$_{\odot }$, respectively. Nevertheless, from a comparison of the lithium
abundances listed in table 2  (at http://cdsweb.u-strasbg.fr/)
 with the standard theoretical predictions, one can
observe a factor of dilution as large as 400, in particular
for stars in the spectral region G8III -- K0III.
Such a contrast between predicted and observed factor of dilution was also
observed by other authors (e.g.: Brown {\it et al.} 1989).
However, the present work which combines a large sample of data from the 
literature (Brown
{\it et al.} 1989; Balachandran 1990; Wallerstein {\it et al.} 1994) shows that 
the dilution 
of lithium
from the spectral type F2III to K5III is far more important than pointed
out up to date. Another important feature that emerges from Figure 
\ref{ALi_BV_Sing} is the 
well known large spread in lithium abundance for a given color index or
spectral type, which could
result from the mixing in masses.
\begin{table}
\caption[]{Statistical results for rotation and lithium abundance by mass 
interval.}
\begin{center}
\begin{tabular}{|l|c|c|c|}
\hline
Mass range   & Correlation  &   rms      & Number   \\
~~($M_\odot$)  & Coefficient  & residual & of stars  \\
\hline
all stars~~           & 0.588~~~~ & 0.711 &  309  \\
$ all (B-V) < 0.7       $ & 0.505~~~~ & 0.588 & 16 \\
$  all (B-V) \geq 0.7   $ & 0.319~~~& 0.658   & 293\\
$ 0.7 < M\leq  1.5 $  & 0.663~~~~ & 0.861 &  73   \\
$ 1.5 < M\leq  2.5 $  & 0.806~~~~ & 0.538 &  91   \\
$ 2.5 < M\leq  3.5 $  & 0.558~~~~ & 0.431 &  110  \\
$  M >  3.5        $  & 0.297~~~~ & 0.688 &  35 \\
\hline
\end{tabular}
\end{center}
\label{tabstat}
\end{table}
A comparison of figures \ref{Vsini_BV_Sing} and \ref{ALi_BV_Sing}
shows, at first glance, no clear link between the drop in rotation near G0III 
and 
the behavior 
of lithium abundance around such  a spectral type. As it has been already outlined, 
rather than
a drop, the lithium abundance shows a gradual decrease. Zahn (1992) and
Pinsonneault {\it et al.} (1989, 1990) have proposed, following different
approaches, that depletion of lithium in single late-type stars is directly
related to the loss of angular momentum. In this context, a correlation
between rotation and lithium content should be expected.
In spite of no clear discontinuity in the lithium contents presented in Figure
\ref{ALi_BV_Sing}, the distribution of lithium abundances represented
by the histogram displayed in Figure \ref{hist_ALi_Single} shows a trend for a
bimodal behavior. In fact, one observes a first mode
around $\log n(Li) = 0.2$, corresponding to stars later than G0III, namely
stars located to the right of the rotational discontinuity, and a second mode
around $\log n(Li) = 3.0$ which is mainly due to the lithium content of
F-type giants. In principle, such a result seems to point in the
direction of a discontinuity in the distribution of lithium at the same spectral
region where the rotational discontinuity is observed, namely around G0III.
One could inquire here about the root cause of this apparent discontinuity in
lithium abundances. Could it be associated to the same root cause
controlling the rotational discontinuity?

In Figure \ref{Vsini_ALi_Single} we show the behavior of lithium abundances as a
function of rotational velocity. Several important features emerge from this 
figure. First of
all, one observes that giant stars presenting the highest lithium content,
typically stars earlier than the spectral type G0III, are also those with
the highest rotation rates. Stars located to the right of the drop in
rotation, namely stars later than G0III, present as a rule, the lowest
rotation rate and lithium content. In fact, these features indicate a trend
leading to a correlation between lithium abundance and rotation along the giant
branch,
which confirms, for single giant stars, the dependence of lithium on rotation, 
observed 
in
 other
luminosity classes.
An additional important feature is the large spread in lithium abundance of the
 slow rotators.
Stars with a $v\sin i$ lower than about 4.0 km\,s$^{-1}$ show a
wide range of lithium abundance values with $\log n(Li)$~ ranging from about
$-1.5$ to the cosmic value, clearly five orders of magnitude.

For a more solid study on the link between rotational velocity
and lithium abundance in single giant stars, we have carried out a least-square 
regression analysis
of the stars listed in Table \ref{tabstat}. A $\log$-linear least-square fit of
$\log n(Li)$ 
against
$v\sin i$ was derived first for the entire sample of stars and then for stars 
located earlier and 
later the spectral type G0III. The least-square
solution  yields very poor linear correlation coefficients and standard 
deviations, as indicated in 
Table \ref{tabstat} and, at first glance, we could regard these results as an 
indication that 
rotation and lithium 
would be poorly correlated in single giant stars.
As a second step, we have segregated the stars by mass intervals.
The stellar masses were estimated from evolutionary tracks computed with
the Tou\-louse-Geneva code for a range of stellar masses between 1 and 4 
M$_{\odot}$ and for
metallicities consistent with population I giants
(see do Nascimento {\it et al.} 2000 for a description).
However, solar composition being relevant
to most objects in the sample, only tracks computed with $[Fe/H]=0$ were 
considered here.
Intrinsic absolute magnitudes M$_{\rm V}$ were
derived from the parallaxes and m$_{\rm V}$ 
magnitudes given by Hipparcos. We have determined the  bolometric corrections 
$BC$
by using the  Buser and Kurucz's relation (1992) between  $BC$ and V-I (again
taken from the Hipparcos Catalogue). The stellar
luminosity and the associated error were computed from the sigma error on the
parallax. The correlation coefficients obtained from this regression analysis,
by interval of mass, are  listed in Table \ref{tabstat},
from which one can observe a clear 
trend for a
linear correlation between lithium and rotation for stars with masses between 
1.5 and 
2.5 solar masses. For the additional mass intervals the results of the present 
least-squares 
solutions indicates a poor correlation between lithium and rotation.

\section{Conclusions}

The distribution of lithium abundance for single giant stars shows a trend for 
a discontinuity near $(B-V)$ = 
0.70, correponding to the spectral type G0III. Such a
discontinuity follows the one observed in rotational velocity. 
In addition, this work shows a clear dependence of lithium abundance on 
rotation, 
in the sense that the highest 
rotators are also the stars presenting the highest lithium content. One 
observes that giant stars, 
presenting the highest lithium content and highest rotation are typically stars 
earlier than the spectral 
type G0III. Stars located to the right of the drop in
rotation, namely stars later than G0III, present as a rule, the lowest
rotation rate and lithium content. In fact, these features show, for 
giant 
stars, the same 
dependence of lithium on rotation observed in other luminosity classes. 
Giant stars with masses $ 1.5 < M/M_{\odot}~\leq  2.5 $  show the more solid 
trend for a 
correlation between 
rotational velocity and lithium abundance. For other mass intervals the 
least-square solutions show a 
trend for poor correlation, indicating that the dependence of lithium content on 
rotation 
is mass dependent.
An additional important feature is the large spread in lithium abundance of the
 slow rotators. Stars with a $v\sin i$ lower than about 4.0 km\,s$^{-1}$ show a
wide range of lithium abundance values with $\log n(Li)$~ ranging from about
$-1.5$ to the cosmic value, namely five orders of magnitude. 
Neverthless, because the sample of giants analyzed in the present work is 
somewhat limited, in particular for late F and early G type stars, additional 
measurements of 
lithium abundance are clearly required for stars located in this spectral 
region, for a 
more solid analysis of this apparent discontinuity in lithium. 

\begin{acknowledgements}

 This work  has been supported by continuous grants from the CNPq Brazilian Agency. J.D.N.Jr. 
acknowledges the CNPq grant 300925/99-9. We are grateful to J. R. P. da Silva for assistance with the 
preparation of the data basis. We would like to thank the referee for his useful comments on the 
manuscprit.      

\end{acknowledgements}

\end{document}